\documentstyle[prl,aps,epsf,multicol]{revtex}
\begin{document}

\title{
Phonon anomalies due to strong electronic correlations
in layered organic metals 
}
\draft
\author{Jaime Merino\cite{email0} and Ross H. McKenzie\cite{email} }
\address{School of Physics, University of New
South Wales, Sydney 2052, Australia}
\date{\today}
\maketitle
\begin{abstract}
We show how the coupling between the phonons and
electrons in 
a strongly correlated metal can result
in phonon frequencies which have a non-monotonic
temperature dependence.
Dynamical mean-field theory is used to study
the Hubbard-Holstein model that describes the
 $\kappa$-(BEDT-TTF)$_2$X family of superconducting
molecular crystals. The crossover with increasing temperature
from a Fermi liquid to a bad metal produces
 phonon anomalies that are consistent
with recent Raman scattering and acoustic experiments.
\end{abstract}

\begin{multicols}{2}
\columnseprule 0pt
\narrowtext

An important problem 
concerning strongly correlated metals
such as heavy fermions, cuprates and organic
superconductors is understanding the
interplay of the strong interactions between
electrons and the interactions between
the electrons and phonons.
This is particularly relevant to understanding
the question of whether the phonons play
any role in superconductivity \cite{gunnarsson}.             
In this Letter we show how 
strong electronic correlations can lead
to phonon frequencies varying with temperature.
Although, our calculations focus on explaining
recent experiments  on a particular family of organic superconductors,
the physics involved is relevant
to other strongly correlated metals in which
there is a significant redistribution of the
electronic spectral weight as the temperature is varied.

The quasi-two-dimensional
organic superconductors, $\kappa$-(BEDT-TTF)$_2$X,
are strongly correlated electron
 systems\cite{kino,kanoda,Ross:98,merino,merino1}.
Recent Raman scattering experiments\cite{lin}
find that in the metallic state
the frequency of certain phonons
associated with the 
BEDT-TTF molecules have a non-monotonic temperature dependence
below 200 K. Acoustic experiments\cite{frikach} 
find that  around 40 K there
is a significant softening of the
speed of longitudinal sound propogating perpendicular 
to the layers.
This softening, of the order of a few per cent,
is more than an order of magnitude larger than the softening
associated with the superconducting transition.
We will show that such a temperature
dependence can arise due to the destruction of 
Fermi liquid quasiparticles that occurs 
above the coherence temperature, $T^*$ \cite{Georges:96,Jarrell,merino}.
Anomalies in acoustic
phonons in the heavy fermion UPt$_3$
have also been seen at temperatures
of the order of $T^*$\cite{muller}.

The simplest possible strongly correlated 
electron model for the                     
$\kappa$-(BEDT-TTF)$_2$X family is
a single band Hubbard model on an anisotropic triangular lattice
at half-filling\cite{kino,Ross:98}.
We have found \cite{merino} that many of the transport
properties of the metallic phase are
consistent with the predictions of
dynamical mean-field theory (DMFT)\cite{Georges:96} which captures
exactly the dynamical fluctuations at each lattice site, but neglects
all non-local spatial correlations.
In order to understand the effect of electronic correlations
on the molecular phonon modes we have
considered the relevant Holstein-Hubbard model
where the phonon amplitude couples 
linearly to the local charge density.
The phonon self-energy
is proportional to the electronic density-density correlation
function which we calculate using DMFT.
When the electron-electron interactions are sufficiently large
we find that
phonon frequencies can have a non-monotonic
temperature dependence.
 This is related to the crossover
from a Fermi liquid to a bad metal that occurs with
Increasing temperature and has a significant effect
on the electronic transport properties\cite{merino}.                       
The actual form of the temperature dependence
of the shift in phonon frequency varies
significantly with the phonon frequency.

We consider the Hubbard-Holstein Hamiltonian
\begin{eqnarray}
H &=& t_1 \sum_{ij,\sigma} (c^\dagger_{i \sigma} c_{j \sigma} + h.c.)
 + t_2 \sum_{ik,\sigma} (c^\dagger_{i \sigma} c_{k \sigma} + h.c.)
\nonumber \\
&+& U \sum_{i} n_{i\uparrow} n_{i\downarrow}- \mu \sum_{i \sigma} n_{i \sigma} 
\nonumber \\
 &+& {g \over \sqrt{2}} \sum_{i \sigma} (a^\dagger_i+ a_i ) n_{i \sigma} +
\omega_{0} \sum_{i} a^{\dagger}_i a_i 
\label{hamilt}
\end{eqnarray}
where the electronic part describes electrons
 on the antibonding orbitals
of each dimer of BEDT-TTF molecules which are located on an 
anisotropic triangular lattice. $t_1$ and $t_2$ are the nearest 
and next-nearest-neighbour
hoppings. 
 $U$ is the Coulomb repulsion between two electrons on the same site
and $\mu$ is the chemical potential. The operator, $c^{\dagger}_i$, creates
an electron on the anti-bonding orbital of the dimer.
The operator, $a^{\dagger}_i$,
creates a phonon at site $i$, which describes a molecular 
vibration of frequency $\omega_0$. We consider                           
only the in-phase 
vibrations of the two phonon modes associated with
the dimer; these can be activated 
by Raman scattering (see below).
$g$ is the coupling between the Raman active phonon
and the electron density on a dimer.

We focus on the parameter regime where the electron-
electron interaction is dominant and we are well away from
any instability  (superconducting or charge-density-wave)
due to the electron-phonon coupling.
We can then decouple the set
of Dyson's equations in the
electron-phonon problem, so that the electron self-energy
contains the electron-electron scattering
mechanism only; any effects coming from the interaction of the 
electrons with phonons on the electron propagator
are neglected.               
The electron Green's function is given by \cite{Georges:96}
\begin{equation}
G_{\sigma}( {\bf k}, i \omega_n ) = {1 \over i \omega_n - \epsilon_{\bf
k} - \Sigma ( i\omega_n) } 
\label{Ge}
\end{equation}
where $\omega_n=  (2n +1) \pi T $ is a Matsubara fermion frequency
for temperature $T$. $\epsilon_{\bf k} = t_1 \cos(k_x) + t_2 \cos(k_x+k_y) $ is the dispersion relation for the anisotropic triangular lattice. 
$\Sigma( i\omega_n)$ is the momentum independent self-energy
computed within DMFT from the associated Anderson impurity problem, 
using the iterative perturbation theory. Details can be found 
elsewhere \cite{Georges:96,merino}.

\begin{figure}
\centerline{\epsfxsize=9cm \epsfbox{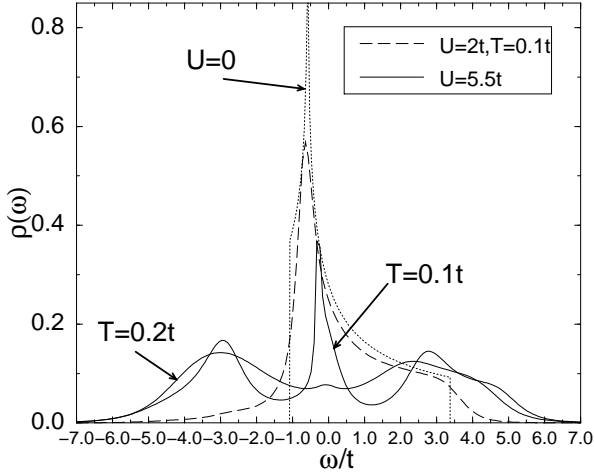}}
\caption{ Spectral densities of
the metallic phase of the Hubbard model at half-filling
calculated for the triangular lattice using
dynamical-mean field theory.
The bare density of states ($U=0$)
is shown as a dotted line.
A peak at the Fermi energy due to a coherent
band of Fermi liquid
quasiparticles and
two incoherent Hubbard bands situated at $\omega = \pm U/2$, are clearly resolved
when the electron-electron
interactions are larger than the metal bandwidth, $W$
($U \ge W = 4.5 t$).
The quasiparticle peak is
strongly temperature dependent and its spectral weight becomes
small as $T > T^*$, $T^*$, being the coherence temperature.
This strong temperature dependence of the spectral density
is the origin of the temperature dependence of the phonon frequencies
discussed in this paper.
}

\label{fig2}
\end{figure}

The phonon problem is solved separately through the associated
Dyson equation
\begin{equation}
D({\bf q}, \omega) = { \omega_0^2 \over  \omega^2 - \omega_0^2 - g^2
\omega_0^2 \Pi({\bf q}, \omega )/2  }
\label{phonon}
\end{equation}
where $\Pi({\bf q}, \omega )$ is the electronic
density-density correlation function (or polarization)
which includes
the full effect of the electron-electron interactions
at the level of DMFT.
For $\Pi({\bf q}, \omega )$ we take the particle-hole
bubble, which, in terms of Matsubara
frequencies is given by
\begin{equation}
\Pi( {\bf q}, i \nu_n) = T 
 \sum_{{\bf k}, \sigma, \omega_n} G_{\sigma}( {\bf k}, i \omega_n )
G_{\sigma}( {\bf k+ q}, i \omega_n + i\nu_n )
\label{bubble}
\end{equation}
and $G_{\sigma}( {\bf k}, i \omega_n )$
 is the Green's function of the electrons 
obtained from the solution of (\ref{Ge}).
The temperature dependence of the polarization predominantly
comes from the temperature dependence of the individual
one-electron Greens functions.
Fig. \ref{fig2} shows how the Fermi
liquid quasiparticle peak in the spectral density of states
is strongly temperature dependent. 
We show results for the case, $t_1=t_2=t$ \cite{note},
which corresponds to a triangular lattice.
      
\begin{figure}
\centerline{\epsfxsize=9cm \epsfbox{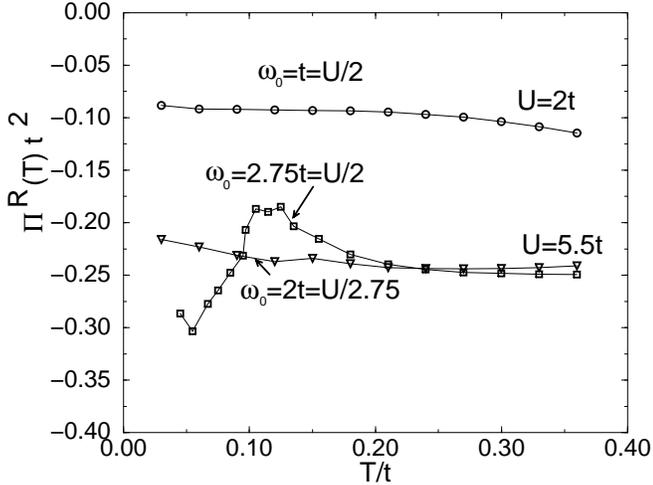}}
\caption{Temperature dependence of the real part of the 
polarization for
different frequencies and correlation strengths.
Due to their coupling to electrons the
phonons undergo a frequency shift that is
proportional to this polarization.
Significant temperature dependence occurs
when two conditions are
simultaneously satisfied: (i) $\omega_0$, is close to $U/2$ and (ii)
electron-electron interactions are sufficiently strong that
there are well-defined Hubbard bands (compare Fig. 1).
For $U/t = 2$ the temperature dependence
is negligible, as expected from the slight
temperature dependence of the spectral density.                            
}
\label{fig3}
\end{figure}

Due to the electron-phonon interactions, the 
phonon frequency is shifted from its bare value $\omega_0$ and 
has a finite lifetime.                       
This shift can be obtained from the poles of 
(\ref{phonon}), which we denote by $\tilde{\omega}_{\bf q}=
\omega_{\bf q}+ i \Gamma_{\bf q}$.
For weak electron-phonon coupling the frequency shift is
\begin{equation}
{\Delta \omega \over \omega_0} \equiv {\omega_{\bf q}-\omega_0 \over \omega_0 }
= {g^2 \over 4} \Pi^R({\bf q}, \omega_0) 
\label{omega}
\end{equation}
where $\Pi^R({\bf q}, \omega_0)$ denotes the real part of the 
polarization.
The phonon damping is  proportional to the imaginary
part of the polarization.
Fig. \ref{fig3} shows the temperature dependence of
the real part of the polarization,  $\Pi^R({\bf q}=0, \omega_0)$,
for different bare phonon frequencies, $\omega_0$.

Our DMFT calculations show that for frequencies close to $ U/2 $
{\it and} sufficiently strong interactions 
the phonon frequency  can have a non-monotonic
temperature dependence near the coherence temperature, $T^*$.
From Fig. \ref{fig3}, we find that $T^* \approx 0.1 t \sim 100 K$, 
for $t \sim 0.1$ eV.
However, this behaviour disappears gradually as $U$ decreases
and becomes smaller than the bandwidth, $W = 4.5t$. This is
clearly seen for $U=2 t$.
The effective mass enhancement
for $U = 5.5 t$ is $m^*/m \approx 3.8$; this is in the 
range of the effective masses found in 
$\kappa$-(BEDT-TTF)$_2$X \cite{merino1}. 
Lin {\it et al.}\cite{lin} have measured 
an anomalous softening below about 100 K of the
Raman frequency shifts in 
the phonons, $\nu_9=$ 505 cm$^{-1}$ and 
$\nu_{60}(B_{3g})=$ 890 cm$^{-1}$ for
$\kappa$-(BEDT-TTF)$_2$ Cu(SCN)$_2$ 
(see Fig. 5 in ref.\cite{lin}).
 The higher frequency mode,    
$\nu_3=$ 1478 cm$^{-1}$, also exhibits
a strong (but monotonic) temperature dependence.
 Table \ref{table1} gives the magnitude of the total
temperature dependence between about 10 and 300 K.
Other phonons do not exhibit such a strong temperature
dependence\cite{eldridge}. Fig. \ref{fig3} and equation (\ref{omega})
implies that
for a coupling $g \sim t$ the temperature dependence can be
as large as 5 \% for $\omega_0 = U/2$
and of the order of 1 \% for larger or smaller frequencies.
The values of $g$ deduced below (see also Table 1)
are consistent with this if $t \sim 0.1 $ eV 
which is also reasonable \cite{merino1}.
However, if $g \sim t \sim \omega_0$
this raises questions about vertex corrections to (\ref{bubble})
and the contribution of electron phonon coupling to
the electronic self energy. Yet the effects shown
in Fig. \ref{fig3} are at most a few per cent
and so we consider that effects that are higher order
in $g$ will be very small.

We find even stronger effects for low-frequency phonons.
 In Fig. 3 the real part of the phonon self-energy\cite{impurit}
 is plotted for ${\bf q}=0$ and
$\omega_0 =0.05 t$.
For strong correlations ($U > W$), a dip in the real part of 
the self-energy appears  at the coherence temperature, $T^*$.
For decreasing values of $U$, the position of the dip moves to higher
temperatures (because $T^*$ increases) and the dip becomes
smaller.  These results could be relevant
to understanding recent acoustic experiments \cite{frikach}.
The velocity of ultrasonic waves which are 
propagating perpendicular to the layers was found
to have a non-monotonic temperature dependence.
These waves have frequencies     
of 100 MHz and velocities of about 2000 cm/s.
The velocity versus temperature shows a broad
dip of a few per cent around 40 K. 
This softening becomes larger 
as the pressure is decreased and
is about three times larger for 
$\kappa$-(BEDT-TTF)$_2$Cu[N(CN)$_2$]Br
than
$\kappa$-(BEDT-TTF)$_2$ Cu(SCN)$_2$.
 Decreasing the pressure or changing
the anion from Cu(SCN)$_2$ to Cu[N(CN)$_2$]Br
corresponds  to 
increasing the electronic correlations or increasing       
of $U/t$. For example, it has been observed that
as the pressure decreases $m^*/m_e$ increases
and the metal-insulator transition is approached \cite{kino,kanoda,Ross:98}.
 Our calculated variation of the position of the dip with $U$
 is in qualitative agreement with the observed
variation of the position of the dip with
pressure (compare Fig. 4 in Ref. \onlinecite{frikach}).
However, caution is in order because further experiments\cite{poirier}
find that the softening only occurs for waves propogating
parallel to the layers when their polarisation is perpendicular
to the layers. Our model can only explain this if modulation of
the interlayer spacing has a much stronger coupling to the electronic charge density
within the dimers than modulation of the interdimer spacing.

\begin{figure}
\centerline{\epsfxsize=9cm \epsfbox{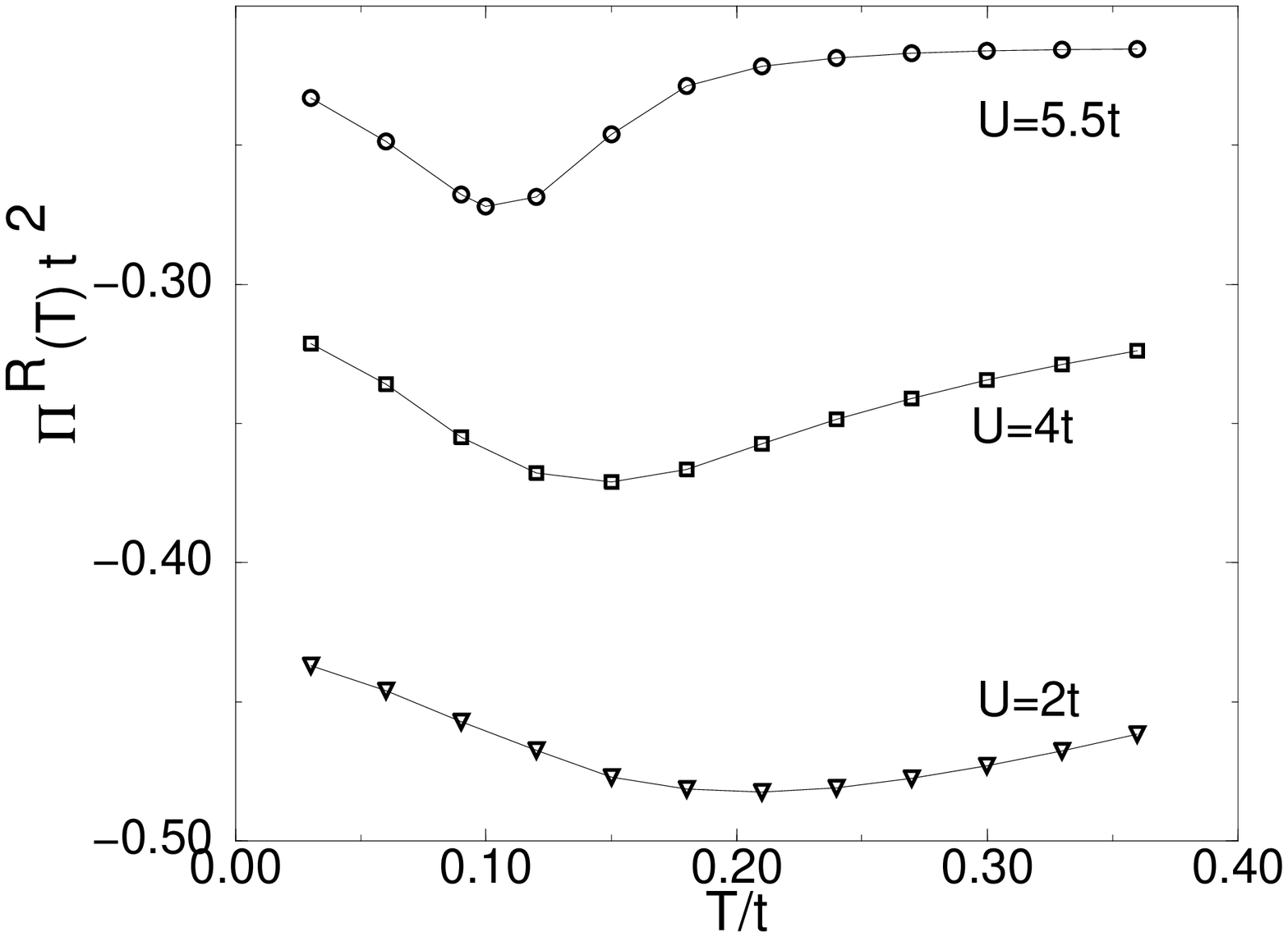}}
\caption{Temperature dependence of the real part of the polarization
for $\omega_0 = 0.05t$ and different values of
$U/t$. A dip appears
at the coherence temperature, $T^*$,
and it becomes more sharply defined as the
electron-electron interactions increase.
The position of the dip shifts towards lower temperatures
as $U/t$ increases.
 }
\label{fig4}
\end{figure}

 The family $\kappa$-(BEDT-TTF)$_2$X
are particulary amenable to study the effect of electronic correlations
on the molecular phonons because
 the dimer structure of the molecular crystal
allows us to extract the electron-phonon
coupling strength $g$ from experimental data.
The crystal structure is such that the BEDT-TTF molecules
are arranged in pairs that are reasonably well
separated from one another. There are three electrons per dimer.
For each phonon mode on a molecule
there is a symmetric and an anti-symmetric combination
on the dimer. By parity conservation,
the anti-symmetric modes are infra-red active
and the symmetric modes are Raman  active.
If there is a Holstein model coupling for
a single molecule the Hamiltonian for a dimer is\cite{rice}
\[
H_{ep}= -t_{0}  \left(c_1^\dagger c_{2} + c_{2}^\dagger c_1 \right)
    + \sum_{i=1}^2  g c_i^\dagger c_i
        \left(a_i + a_i^\dagger \right)
   + \omega_0  a_i^\dagger a_i
\]
where $t_0$ is the transfer integral for hopping
between molecules {\it within} the dimer.
This is much larger than the hoppings $t_1$ and
$t_2$ {\it between}
dimers that appears in (\ref{hamilt}) \cite{Ross:98}.
The bonding and anti-bonding orbitals associated
with the dimer are separated by $2 t_{0} $,
even in the presence of a Hubbard term.
The Hamiltonian can be re-written as
\[
H_{ep}= -t_{0}  \left(c_1^\dagger c_{2} + c_{2}^\dagger c_1 \right)
    + \sum_{\alpha=\pm}  { g \over \sqrt{2} } n_\alpha
        \left(a_\alpha + a_\alpha^\dagger \right)
   + \omega_0  a_\alpha^\dagger a_\alpha
\]
where $a_\pm = {1 \over \sqrt{2} } (a_1 \pm a_2 )$
and $n_\pm = c_1^\dagger c_1 \pm c_2^\dagger c_2$.
The $n_+$ term is the same as the number operator, $n_i$, appearing 
in (\ref{hamilt}). It is the fluctuations in this term that produce
the temperature dependence shown in Figures 
\ref{fig3} and \ref{fig4}.
In contrast, the infrared active mode
will not couple to these fluctuations and so
should have no such temperature dependence.

If the number of electrons on the dimer is fixed then
the symmetric mode has frequency $\omega_0$.
A second-order perturbative calculation\cite{rice}
predicts that for $\omega_0 \ll t_{0}$,
the infra-red frequency $\omega_{IR}$ will be smaller than
the Raman frequency ($\omega_{R}=\omega_0$),
\begin{equation}
{\omega_{IR} - \omega_R \over \omega_R} = - {g^2 \over  2 \omega_0 t_{0}}.
\label{shift}
\end{equation}
Table I lists the details of the measured frequency shifts for
three different modes
for $\kappa$-(BEDT-TTF)$_2$Cu(SCN)$_2$.
For $t_{0} = 0.3 $ eV \cite{Fortunelli} these values are
used together with (\ref{shift}) to evaluate
the electron-phonon coupling $g$.
The values obtained are in reasonable agreement
with those found in a
MNDO frozen phonon quantum chemistry calculation\cite{shumway}
on a single BEDT-TTF molecule.

Zeyher and Zwicknagl\cite{zeyher}
considered the frequency shift of phonons when they
enter the superconducting phase. Within the framework
of BCS theory, phonons with frequency $\omega_0$ much
smaller than the Fermi energy and much
larger than the superconducting gap $\Delta$
will harden by an amount
\begin{equation}
{\omega_s - \omega_R \over \omega_0} =  {8 g^2 N(0) \over  \omega_0}
\left( { \Delta \over \omega_0 } \right)^2 \ln \left( { \omega_0 \over
\Delta} \right)
\label{shift-sc}
\end{equation}
where $N(0)$ is the density of states per spin
at the Fermi energy.
There should be no shift in the frequency of
the infra-red active mode on entering the
superconducting phase.
Using $\Delta \sim 2 k_B T_c \sim 0.002$ eV
and the values for $g$ in Table I we have
evaluated the estimated shift in the phonon
frequency for three different modes in Table \ref{table1}.
Eldridge {\it et al}.\cite{eldridge} found that
in $\kappa$-(BEDT-TTF)$_2$Cu[N(CN)$_2$]Br
the $\nu_{60}(B_{3g})$ mode at 890 cm$^{-1}$
hardened by about 0.2 \% on
entering the superconducting phase.
A shift in this mode was not detected in
$\kappa$-(BEDT-TTF)$_2$Cu(SCN)$_2$.
No frequency shift in the other
high-frequency modes was detected to within about 0.1 \%.
However, one should be cautious about making quantitative
comparisons between experiment and Table I
since (\ref{shift-sc}) is only valid for $\omega_0 \ll t$
and we have $\omega_0 \sim t$.
Pedron {\it et al.} did observe the hardening of
modes with frequencies ranging from 27 to 134 cm$^{-1}$\cite{pedron}.
For acoustic phonons the softening will scale
like $(\omega_0/\Delta)^2$ and so be much
smaller than the effects shown in Fig. 
\ref{fig4}.

In conclusion, we have shown how in a strongly
correlated metal the redistribution of 
spectral weight over the scale of the band width with varying temperature
can result in phonon frequencies with anomalous
temperature dependence. The effects involved are
larger than those associated with the superconducting
transition because the latter only involves a
redistribution of spectral weight over energies
of the order of the superconducting gap
which is much less than the band width.

This work was supported by the Australian Research
Council. We thank J. E. Eldridge, J. B. Marston
and M. Poirier for helpful discussions.

\newpage
{\onecolumn
\vskip0.1cm
\begin{table}
\caption{Parameters for three molecular phonons in
$\kappa$-(BEDT-TTF)$_2$Cu(SCN)$_2$.
The experimental values are taken from Reference \protect\onlinecite{eldridge2}
The difference between the infra-red frequency, $\omega_{IR}$
and the  Raman frequency, $\omega_{R}$, and equation (\protect\ref{shift})
is used to estimate (est.) the electron-phonon couplings $g$.
These values are  compared with values of $g$ calculated
by a quantum chemistry (QC) calculation \protect\cite{shumway}.
$\Delta \omega_R(T)$ is the observed
magnitude of the temperature dependence of the
Raman-active mode between about 10 K and 300 K. $\Delta \omega_{SC}$ is
 an estimate of the increase of the phonon frequency
on entering the superconducting phase, based on equation (\ref{shift-sc}).
}
\vskip0.5cm
\label{table1}
\begin{tabular}{dddddddd}
  &\multicolumn{2}{c}
{Observed Frequency (cm$^{-1}$)} & &
\multicolumn{2}{c} {e-p coupling, $g$ (eV)}  & & \\
Mode& Infrared($\omega_{IR}$) & Raman($\omega_R$) &
${ \omega_R-\omega_{IR} \over \omega_R} (\%)$ & est.  &
 QC    & ${\Delta \omega_R(T) \over \omega_R }(\%)$ &
${\Delta \omega_{SC} \over \omega_0 } (\%)$  \\
\hline
$\nu_3$  &  1276  &  1478 &   14   & 0.14 &  0.14 & 0.3  & 0.13 \\
$\nu_9$  &  431  &  505 &   15   & 0.07 &  0.08 & 1  &  0.7 \\
$\nu_{60}$  &  881  &  890 &   1   & 0.02 & ?  & 0.3  & 0.02  \\
\end{tabular}
\end{table}
}

\twocolumn

\end{multicols}

\begin{references}

\bibitem[*]{email0}e-mail: merino@phys.unsw.edu.au
\bibitem[**]{email}New address: Department of Physics,
University of Queensland, Brisbane,  4072, Australia;
e-mail: mckenzie@physics.uq.edu.au

\bibitem{gunnarsson} O. Gunnarsson, Rev. Mod. Phys.
 {\bf 69}, 575 (1997).

\bibitem{kino}
H. Kino and H. Fukuyama,
J. Phys. Soc. Jap.  {\bf 65}, 2158 (1996).

\bibitem{kanoda}
K. Kanoda,
 Physica C {\bf 282}, 299 (1997).

\bibitem{Ross:98} R. H. McKenzie,
Comments Cond. Mat. Phys. {\bf 18}, 309 (1998).

\bibitem{merino} J. Merino and R. H. McKenzie, Phys. Rev. B {\bf 61},
7996 (2000).

\bibitem{merino1} J. Merino and R. H. McKenzie, Phys. Rev. B,
to appear 15 July (2000).

\bibitem{lin} Y. Lin {\it et al}.,
Phys. Rev. B {\bf 58} R599 (1998).

\bibitem{frikach} K. Frikach {\it et al}.,
Phys. Rev. B {\bf 61}, R6491 (2000).

\bibitem{Georges:96} A. Georges {\it et al}.,
Rev. Mod. Phys. {\bf 68}, 13 (1996).

\bibitem{Jarrell}
Th. Pruschke, M. Jarrell, and J. K. Freericks,
 Adv.  Phys.  {\bf 44}, 187   (1995).

\bibitem{muller}
V. M\"uller, {\it et al}.,
Phys. Rev. Lett. {\bf 56}, 248 (1986);
J. Grolle, V. M\"uller, and K. H. Bennemann,
Phys. Rev. B {\bf 35}, 4493  (1987).
Both  the temperature dependence
of the acoustic attenuation and the thermopower
exhibit peaks around 10 K.

\bibitem{note}
Relaxing this condition produces only
small quantitative changes in our results.

\bibitem{eldridge} J. E. Eldridge {\it et al}.,
Phys. Rev. B {\bf 57}, 597 (1998).

\bibitem{impurit}
In the calculations shown in Fig. \ref{fig4} a small
imaginary part has been added to the self-energy
($\Sigma^I(\omega=0)=0.1t$)
 to allow for the effect of impurities and to be consistent
with the observed resistivity at low temperatures
[K. Murata {\it et al.}, Synth. Met. {\bf 27} A623, (1988)].

\bibitem{poirier}
M. Poirier, 
private communication.

\bibitem{rice}
M. J. Rice,
Solid State Commun. {\bf 31}, 93  (1979).
Note, that Rice's calculations are
for two electrons per dimer. For the case of
three electrons per dimer considered here
the Hubbard $U$ has no effect on the phonon
frequency shift.

\bibitem{Fortunelli} A. Fortunelli and A. Painelli, J. Chem. Phys. {\bf
106}, 8051 (1997).



\bibitem{shumway}
J. Shumway, S. Chattopadhyay and S. Satpathy,
Phys. Rev. B {\bf 53}, 6677  (1996).

\bibitem{zeyher}
R. Zeyher and G. Zwicknagl,
Z. Phys. B {\bf 78}, 175 (1990).

\bibitem{pedron}
D. Pedron  {\it et al.}, Synth. Met. {\bf 85}, 1509 (1997).

\bibitem{eldridge2}
 J. E. Eldridge {\it et al}., Spectrochim.
Acta A  {\bf 53}, 565 (1997).

\end{references}
\end{document}